\documentclass[twocolumn,preprintnumbers,amsmath,amssymb,floatfix,prd]{revtex4}

\usepackage{graphicx}
\usepackage{dcolumn}
\usepackage{bm}

\begin{document}

\title{Study of the angular coefficients and corresponding helicity cross sections \\ of the $\bm{W}$ boson in hadron collisions}

\author{John Strologas}%
\affiliation{%
Department of Physics and Astronomy, University of New Mexico, Albuquerque, NM 87131}%
\author{Steven Errede}
\affiliation{%
Department of Physics, University of Illinois at Urbana-Champaign, Urbana, IL 61801}%

\date{March 30, 2005}

\begin{abstract}
We present the Standard Model prediction for the eight angular coefficients
of the $W$ boson, which completely describe its differential
cross section in hadron collisions.
These coefficients are ratios of the $W$ helicity
cross sections and the total unpolarized cross section.  We also suggest
a technique to experimentally extract the coefficients.
\end{abstract}


\maketitle
\section{\label{sec:1} Introduction}
The analytical study of the $W$ boson is essential for the
understanding of many open questions related to the electroweak
physics, like the origin of the electroweak symmetry breaking and
the source of the $CP$ violation.  Since its discovery the $W$
hadronic cross section, mass, and width have been measured with
great precision \cite{pdg}.  On the other hand complete physical
information is contained in the boson's angular distribution in
three dimensions, given by its differential cross section, which 
can be written as a sum of helicity cross sections.
These quantities are related to the nature of the electroweak processes, 
the $W$ polarization, and the presence of QCD effects.  In this paper we
address these issues in the case of $W$ produced in hadron
collisions.

 The total differential cross section of the $W$ production in
a hadron collider \cite{mirkesnpb} is given by the equation:
\begin{eqnarray}
\frac{d{\sigma}}{dq_T^2dyd\cos{\theta}d\phi} &=&
\frac{3}{16\pi}\frac{d\sigma^u}{dq_T^2dy}[(1+\cos^2{\theta})  \nonumber \\
&+& \frac{1}{2}A_0(1-3\cos^2{\theta})
+ A_1\sin{2\theta}\cos{\phi} \nonumber \\
&+&\frac{1}{2}A_2\sin^2{\theta}\cos{2\phi}
+ A_3\sin{\theta}\cos{\phi} \nonumber \\
&+& A_4\cos{\theta} +A_5 \sin^2 \theta \sin 2\phi  \nonumber \\
&+& A_6\sin{2\theta}\sin{\phi}
+ A_7\sin{\theta}\sin{\phi}]
\label{eq1}
\end{eqnarray}
where $q_T$ and $y$ are the transverse momentum and the rapidity
of the $W$ in the lab frame and $\theta$ and $\phi$ are the polar
and azimuthal angles of the charged lepton from the $W$ decay in
the Collins-Soper (CS) frame \cite{cs}. The CS frame is used
because in this frame we can experimentally reconstruct the
azimuthal angle $\phi$ and the polar quantity $|\cos{\theta}|$.
Our ignorance of the $W$ longitudinal momentum, which is due to
our inability to measure the longitudinal momentum of the
neutrino in a hadron collider, introduces a two-fold ambiguity on the sign of
$\cos\theta$.  The quantity $d\sigma^u/dq_T^2dy$ is the angles-integrated 
unpolarized cross section.

The dependence of the cross section on the leptonic variables
$\theta$ and $\phi$ is completely manifest and the dependence on
the hadronic variables $q_T^2$ and $y$ is completely hidden in
the angular coefficients $A_i(q_T,y)$. This allows us to treat the
problem in a model independent manner since all the hadronic
physics is described implicitly by the angular coefficients and it
is decoupled from the well understood leptonic physics. The angular
coefficients are ratios of the helicity cross sections of the $W$
and $d\sigma^u/dq_T^2dy$. In order to explicitly separate the hadronic
from the leptonic variables, the helicity amplitudes were used to
describe the hadronic tensor associated with the hadronic
production of the $W$ \cite{lam}.  The leptonic tensor on the
other hand is analytically known, leading to analytic functions of
the angles of the charged lepton in Equation (\ref{eq1}). It is
common to integrate Equation (\ref{eq1}) over $y$ and study the
variation of the angular coefficients as a function of $q_T$.

If the $W$ is produced with no transverse momentum, it is
polarized along the beam axis because of the V-A nature of the
weak interactions and helicity conservation. In that case $A_4$ is
the only non-zero coefficient. If only valence quarks contributed
to the $W^{\mp}$ production, $A_4$ would equal 2, and the angular
distribution (\ref{eq1}) would be $\sim (1\pm\cos\theta)^2$, a
result that was first verified by the UA1 experiment \cite{ua1}.

If the $W$ is produced with non-negligible transverse momentum,
balanced by the associated production of jets, the rest of the
angular coefficients are present and the cross section depends on
the azimuthal angle $\phi$ as well.  The last three angular
coefficients -- $A_5$, $A_6$, and $A_7$ -- are non-zero only if gluon
loops are present in the production of the $W$ \cite{hagiwara}.
Hence, in order to study all the angular coefficients and associated
helicity cross sections of the $W$ in a hadron collider, we have
to consider the production of the $W$ with QCD effects at least up
to order $\alpha_s^2$.

The importance of the determination of the $W$ angular
coefficients is discussed in \cite{mirkesprd}, and is summarized
here. This study allows us to measure for the first time the differential
cross section of the $W$ and study its polarization, since the
angular coefficients are related to the helicity cross sections.
It also helps us verify the QCD effects in the production of the
$W$ up to order $\alpha_s^2$. In addition, $A_3$ is only affected
by the gluon-quark interaction and its measurement could constrain 
the gluon parton distribution functions. Moreover,
the next-to-leading order (NLO)
coefficients $A_5$, $A_6$, and $A_7$ are $P$-odd and $T$-odd and may play an
important role in direct $CP$ violation effects in $W$ production
and decay \cite{hagiwara}. Finally, a quantitative understanding of the
$W$ angular distribution could be used to test new theoretical
models and to facilitate new discoveries.

In this paper we present the Standard Model prediction for the
angular coefficients $A_i$ as a function of $q_T$, for
proton-antiproton collisions at 1.8 TeV, using the DYRAD Monte
Carlo program \cite{giele}, an event generator of $W$+jet up to
order $\alpha_s^2$.  Three of the coefficients, including $A_4$
which is the dominant one up to $q_T=100$ GeV, are presented for
the first time. 
We also suggest a new method for the extraction
of the angular coefficients.

\section{\label{sec:2} \boldmath Standard Model Prediction for the $W$ Angular Coefficients}
\begin{figure}
\includegraphics[scale=.44]{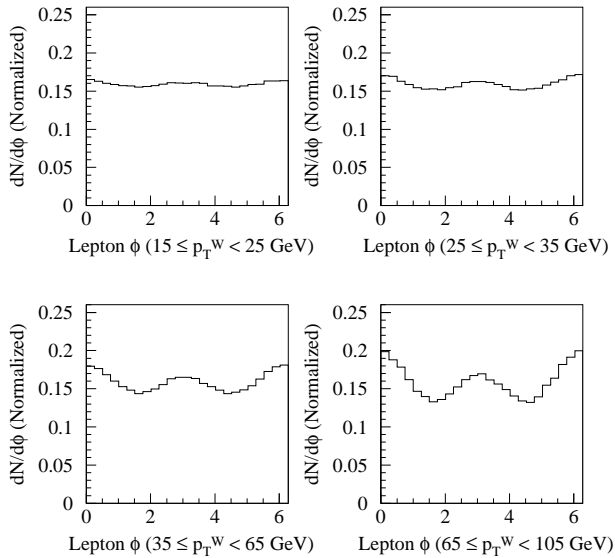}
\caption{The charged lepton $\phi$ distribution in the Collins-Soper $W$ rest-frame for four $q_T$ regions.  The distributions are normalized to unity.}
\label{f1}
\end{figure}
\begin{figure}
\includegraphics[scale=.44]{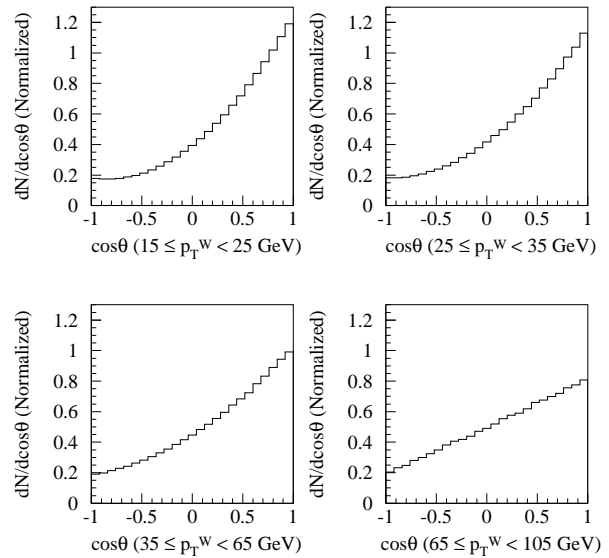}
\caption{The charged lepton $\cos \theta$ distribution in the Collins-Soper $W$ rest-frame for four $q_T$ regions.  The distributions are normalized to unity.}
\label{f2}
\end{figure}

In order to study the angular distribution of the $W$ we have to choose a
particular charge for the boson.  In this paper we present the results for the
$W^-$.  The angular coefficients for the $W^+$ can be extracted by
$CP$ transformation.  In the Collins-Soper frame, the $CP$
transformation leaves $\phi$ unchanged and takes $\theta$ to
$\pi-\theta$. If we assume that Equation (\ref{eq1}) describes $W^-$
bosons, we have to change the sign of coefficients $A_1$, $A_4$,
and $A_6$, in order to describe $W^+$ bosons, without changing the
definition of the Collins-Soper frame.

We generate Monte Carlo $W$+jet events up to $\alpha_s^2$, including up to one
gluon loop, from proton-antiproton collisions at $\sqrt{s}=$ 1.8
TeV, using the DYRAD generator.  We run with minimal kinematic and
acceptance cuts, with a minimum transverse energy for the jet of 10 GeV and jet-jet 
angular separation of $\Delta R = \sqrt{(\Delta \eta_{\rm lab})^2+(\Delta\phi_{\rm lab})^2} = 0.7$, 
in the pseudorapidity-phi space in the lab frame.  The CTEQ4M
parton distribution functions were used.

We measure the $\theta$ and $\phi$ angles of the charged lepton in
the Collins-Soper $W$ rest-frame. At the Monte Carlo event-generator 
level, we know the momentum of the neutrino, so there is no two-fold ambiguity on
the sign of $\cos\theta$. The CS frame is the rest-frame of the
$W$ where the $z$-axis bisects the angle between the proton momentum
($\roarrow{p}_{\rm{CS}}$) and the opposite of the antiproton
momentum ($-\roarrow{\bar{p}}_{\rm{CS}}$) in the CS frame. 
The signs of the angular coefficients depend on the way the
Collins-Soper $x$-axis and $y$-axis are defined.  In this paper, we define them so
that the x-z plane coincides with the $p_{\rm{CS}}-\bar{p}_{\rm{CS}}$
plane and the positive $y$-axis has the same direction as
$\roarrow{p}_{\rm{CS}} \times \roarrow{\bar{p}}_{\rm{CS}}$. The SM
distribution of the $\phi$ and $\cos\theta$ for four $q_T$ bins
(15-25, 25-35, 35-65 and 65-105 GeV) is shown in Figures
\ref{f1} and \ref{f2} respectively.
We note that at low transverse momentum of the $W$, the $\phi$ distribution
is almost flat, whereas the $\cos \theta$ distribution almost follows the
$(1+\cos\theta)^2$ law.  In the fourth $q_T$ bin, there is
a strong $\phi$-dependence of the cross section and the
$\cos \theta$ distribution is almost a straight line ($|\cos\theta|$
is flat).  There is a correlation between $\cos\theta$ and $M_T^W$,
with low $\cos\theta$ corresponding to low $M_T^W$ events.

To calculate the
angular coefficients from the angles of the charged lepton,
we use the method of moments.
We first define the {\it moment} of a function $m(\theta,\phi)$ as 
\begin{equation}
<m(\theta,\phi)>=\frac{\int\int d\sigma(q_T,y,\theta,\phi)m(\theta,\phi)d\cos{\theta}d\phi}
{\int\int d\sigma(q_T,y,\theta,\phi)d\cos{\theta}d\phi}
\label{mom}
\end{equation}
\begin{figure}[t]
\includegraphics[scale=.41]{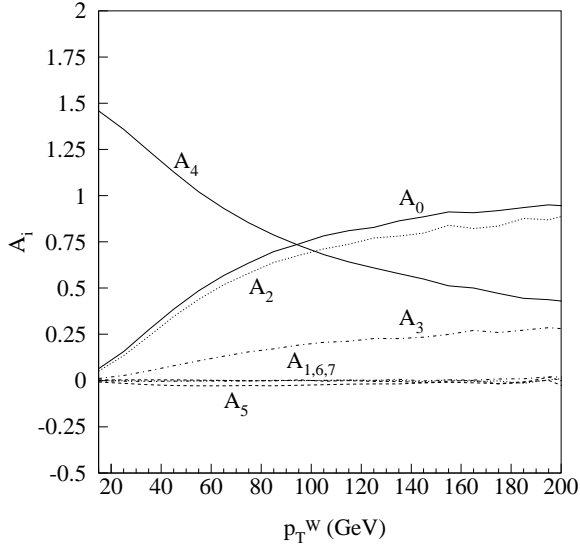}
\caption{The Standard Model prediction for the angular
coefficients of the $W$ produced in a collider at $\sqrt{s}$=1.8
TeV. QCD effects are included up to order $\alpha_s^2$.}
\label{f3}
\end{figure}
\begin{figure}[h]
\includegraphics[scale=.41]{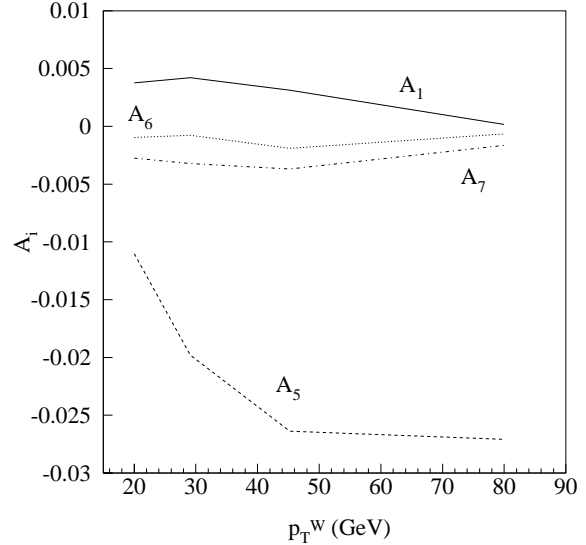}
\caption{The Standard Model prediction for the next-to-leading order coefficients ($A_5$, $A_6$, and $A_7$) and $A_1$ for the $W$ production in a collider at $\sqrt{s}$=1.8 TeV.  The coefficients $A_5$, $A_6$, and $A_7$ are present
only if gluon loops are included in the calculation and they are
$T$-odd and $P$-odd.}
\label{f4}
\end{figure}
\begin{figure}
\includegraphics[scale=.41]{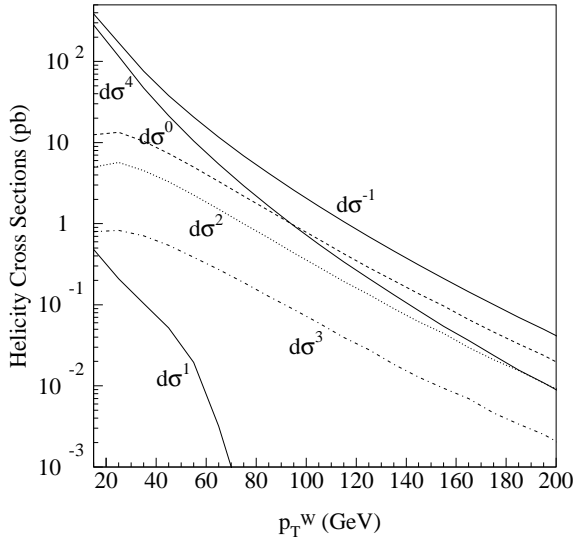}
\caption{The Standard Model prediction for the helicity cross
sections of the $W$ produced in a collider at $\sqrt{s}$=1.8 TeV.
QCD effects are included up to order $\alpha_s^2$.} \label{f5}
\end{figure}
We can easily prove that:
\begin{eqnarray}
<m_0>&\equiv&<\frac{1}{2}(1-3\cos^2{\theta})>=\frac{3}{20}(A_0-\frac{2}{3})\nonumber \\
<m_1>&\equiv&<\sin{2\theta}\cos{\phi}>=\frac{1}{5}A_1\nonumber \\
<m_2>&\equiv&<\sin^2{\theta}\cos{2\phi}>=\frac{1}{10}A_2\nonumber \\
<m_3>&\equiv&<\sin{\theta}\cos{\phi}>=\frac{1}{4}A_3\nonumber \\
<m_4>&\equiv&<\cos{\theta}>=\frac{1}{4}A_4\nonumber \\
<m_5>&\equiv&<\sin^2{\theta}\sin{2\phi}>=\frac{1}{5}A_5\nonumber \\
<m_6>&\equiv&<\sin{2\theta}\sin{\phi}>=\frac{1}{5}A_6  \nonumber \\
<m_7>&\equiv&<\sin{\theta}\sin{\phi}>=\frac{1}{4}A_7 \nonumber \\
\label{big}
\end{eqnarray}

For a set of discrete generator (or experimental) data, we
substitute the integrals of Equation (\ref{mom}) by sums and the
cross section values by the weights $w_i$ of the Monte Carlo events.
\begin{equation}
<m(\theta,\phi)>=\frac{\sum_{i=1}^{N}m(\theta_i,\phi_i)w_i}
{\sum_{i=1}^{N}w_i}
\label{mom2}
\end{equation}

By solving Equations (\ref{big})
for the angular coefficients and substituting the moments
by the discrete expressions (\ref{mom2}),
we extract the Standard Model prediction.
By ignoring the $W$ rapidity , we actually calculate the
$y$-integrated angular coefficients, which are now functions
of just $q_T$.  The results are shown in Figure \ref{f3}.
The angular coefficients $A_1$, $A_4$, and $A_6$ are presented
for the first time.
The DYRAD Monte Carlo generator is more reliable for
$q_T>10$ GeV, which is also the transverse energy cut for
our jets, and this value determines the minimum of our $q_T$-axis.
The maximum is determined by the Monte Carlo statistics.

We notice that indeed $A_4$ is the only surviving major coefficient at
low $q_T$ values. It is also the only angular coefficient that decreases as
$q_T$ increases.   The angular coefficient $A_1$, although it is a
leading order (LO) coefficient, is much smaller than the leading order
coefficients ($A_{0}$, $A_{2}$, $A_{3}$, and $A_{4}$) and
comparable to the next-to-leading order ones ($A_{5}$, $A_{6}$,
and $A_{7}$).  
The coefficients $A_0$ and $A_2$ would be exactly
equal if gluon loops were not included \cite{mirkesnpb}. At order $\alpha_s^2$,
$A_0$ is consistently greater than $A_2$. To better determine
$A_1$, $A_5$, $A_6$, and $A_7$ we use the four $q_T$ bins, to improve the
statistics. The result is shown in Figure \ref{f4}. There are
relations that directly connect the angular coefficients with the
helicity cross sections of the $W$ \cite{mirkesnpb}. We first
extract the unpolarized cross section of the $W$ as a function of
$q_T$ and using our prediction for the angular coefficients, we
arrive at the Standard Model prediction for the $W$ helicity cross
sections at $\sqrt s = 1.8$ TeV, shown in Figure \ref{f5}. Here
$d\sigma_i$ is the helicity cross section that corresponds to the
angular coefficient $A_i$.
\section{\boldmath Experimental Determination of the $W$ Angular Distribution}
E. Mirkes \cite{mirkesnpb} first realized the problem of directly measuring
the angular coefficients.  The angular distributions of Figures
\ref{f1} and \ref{f2} are seriously distorted
after the effects of the detector are considered and quality
cuts are imposed on the data sample.
\begin{figure}
\includegraphics[scale=.44]{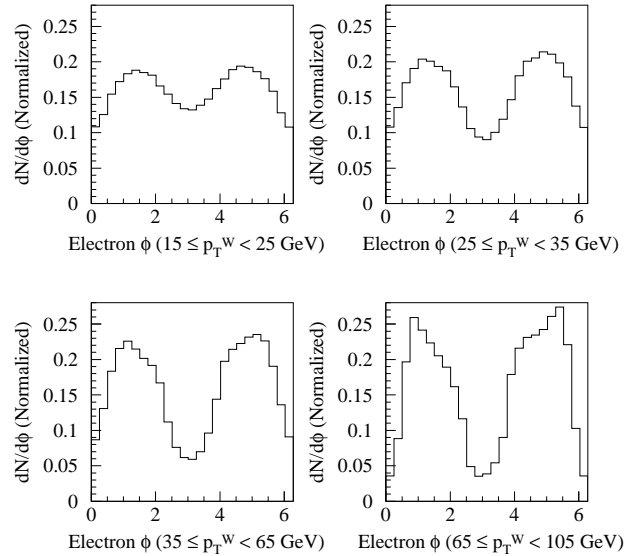}
\caption{The experimentally observed electron $\phi$ distribution in the Collins-Soper $W$ rest-frame for four $q_T$ regions.  The distributions are normalized to unity.}
\label{f6}
\end{figure}
\begin{figure}
\includegraphics[scale=.44]{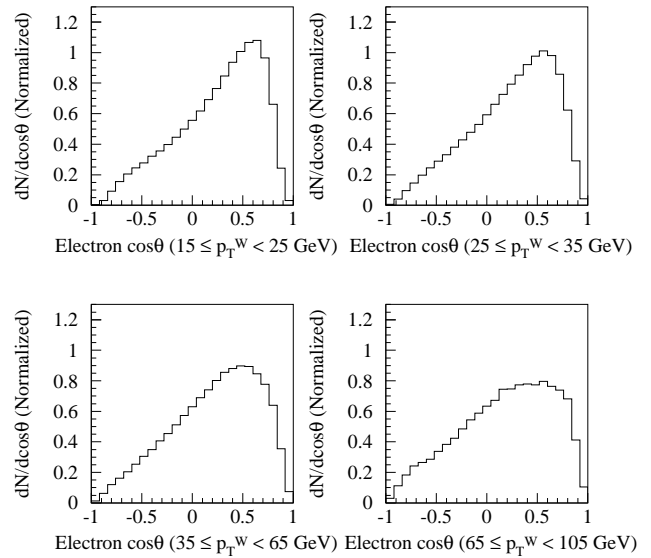}
\caption{The experimentally observed muon $\cos \theta$ distribution in the Collins-Soper $W$ rest-frame for four $q_T$ regions.  The distributions are normalized to unity.}
\label{f7}
\end{figure}
\begin{figure}
\includegraphics[scale=.44]{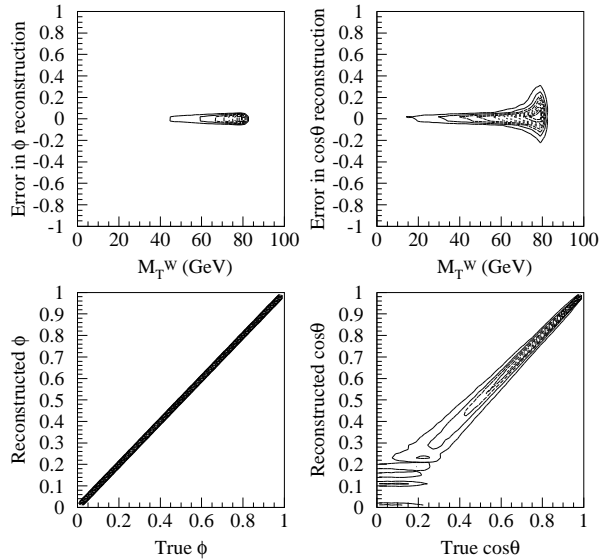}
\caption{The error in $\phi$ reconstruction due to the selection
of the $W$ mass is considerably lower than the error in
$\cos\theta$ reconstruction, especially at high $M_T^W$ values
(upper plots).  This corresponds to better reconstruction of the
$\phi$, while the $\cos\theta$ reconstruction is worse, especially
at low values of $|\cos\theta|$ (lower plots).} \label{f8}
\end{figure}
To study the effect, we treat the generator leptons as electrons
and we pass them through a detector simulator \cite{footnote}. The new $\phi$ and
$\cos\theta$ distributions are shown in Figures \ref{f6} and
\ref{f7} respectively. The shapes of the muon distributions are the
same as that for electron $W$+jet events, however less events are detected, 
because of the lower muon
acceptance of a typical hadron collider detector. The main reason
for the difference between Figures \ref{f1}, \ref{f2} and Figures
\ref{f6}, \ref{f7} is the leptons transverse momentum cuts
($p_T^l>20$ GeV and $p_T^{\nu}>20$ GeV) and the charged lepton
rapidity cut (central leptons are considered, $|y^l|<1$).

The problem of distortion of the $\phi$ and $\cos\theta$ distributions
due the detector effects and quality cuts is not the only one.
A more fundamental problem is the actual measurement of these
angles.  To measure them, we need to reconstruct the $W$ in
the three dimensional momentum space, in order to boost to
its center of mass.  The longitudinal momentum of the neutrino
is not measured, but it is constrained by the mass of the $W$,
based on equation:
\begin{equation}
p_z^{\nu} = \frac{1}{(2p_T^l)^2}\left (Ap_z^l \pm
E^l\sqrt{A^2-4(p_T^l)^2(p_T^{\nu})^2}\right )\label{pznu}
\end{equation}
where
\begin{equation}
A = M_W^2 +q_T^2-(p_T^l)^2-(p_T^\nu)^2, \nonumber
\end{equation}
$E^l$ is the energy of the charged lepton, $p_T^l$ and $p_T^{\nu}$
are the transverse momentum of the charged lepton and the neutrino
and $p_z^l$ is the longitudinal momentum of the charged lepton.
The two solutions for the longitudinal momentum of
the neutrino lead to two solutions for the $W$ longitudinal
momentum. Both solutions correspond to the same $\phi$ but to
opposite $\cos\theta$ values.

Moreover, according to Equation (\ref{pznu}), for each event, we
have to input a mass for the $W$ to get $p_Z^{\nu}$ and eventually
$\phi$ and $|\cos\theta|$. The mass of $W$ is not known on
event-by-event basis, we just know its pole mass and its
Breit-Wigner width. Based on these two established values, we can
plot the uncertainty in the measurement of $\phi$ and
$|\cos\theta|$ introduced by the uncertainty in the mass of the
$W$.  For each Monte Carlo event we generate $W$ masses that are
greater than the transverse mass for the particular event and
follow the Breit-Wigner distribution.  In Figure \ref{f8} we see
that the systematic error on the measurement of $\phi$ is very
small, but the $\cos\theta$ systematic error is significant,
especially at low $|\cos\theta|$ and at big values of the
transverse mass of the $W$.  This makes the direct measurement of
the $|\cos\theta|$ distribution more challenging.
\section{Experimental Extraction of the Angular coefficients}

In \cite{mirkesprd} it is suggested that the experimental
distributions of Figures \ref{f6} and \ref{f7} should be divided
by the Monte Carlo distributions obtained using isotropic $W$
decays.  This method results in distributions similar to those
shown in Figures \ref{f1} and \ref{f2} and the extraction of the
angular coefficients is easier. Here we present a method that does not
bias the experimental data by Monte Carlo data.  Instead, it uses
the knowledge of the detector and its effect on the theoretical
distributions.  We will demonstrate the method for the $\phi$
analysis.

If we integrate Equation (\ref{eq1}) over $\cos\theta$ and $y$, we
get:
\begin{eqnarray}
\frac{d\sigma}{dq_T^2d\phi}=C'(1+\beta_1\cos{\phi}+\beta_2\cos{2\phi} \nonumber \\
+\beta_3\sin{\phi}+\beta_4\sin{2\phi})
\label{eqphi}
\end{eqnarray}
where
\begin{eqnarray}
C'=\frac{1}{2\pi}\frac{d\sigma}{dq_T^2},\: \beta_1=\frac{3\pi}{16}A_3,\:\beta_2=\frac{A_2}{4}  \nonumber \\
\beta_3=\frac{3\pi}{16}A_7,\:\beta_4=\frac{A_5}{2}
\end{eqnarray}
The observed $\phi$ distribution is given by Equation
(\ref{eqphi}), only if we ignore the effects of the detector and
kinematic cuts.
In any other case, there is an acceptance and
efficiency function $ae(q_T,\cos\theta,\phi)$ which multiplies
(\ref{eq1}) before it is integrated over $\cos\theta$ and as a
result, no angular coefficient is completely integrated out. In
the actual data, what we measure is the number of events, which is:
\begin{eqnarray}
N(q_T,\phi)&=& \int \frac{d\sigma}{d q_T d\phi d\cos\theta} ae(q_T,\cos\theta,\phi)
d\cos\theta \int {\cal L} dt \nonumber \\
&+& N_{bg}(q_T,\phi)
\label{new}
\end{eqnarray}
where ${\cal L}$ is the luminosity and $ae(q_T,\cos\theta,\phi)$ is the
acceptances and efficiencies for the particular $W$ transverse
momentum and pixel in the
$(\cos\theta,\phi)$ phase space.  $N_{bg}(q_T,\phi)$ is the background for
the given $\phi$ bin and $q_T$.
If we combine Equations (\ref{new}) and (\ref{eq1}), then the
measured distribution is:
\begin{equation}
N(q_T,\phi)= C' (f_{-1} + \sum_{i=0}^{7} A_i f_i) + N_{bg}(q_T,\phi)
\label{new2}
\end{equation}
where $C' = C\int {\cal L} dt$, and $f_i$ are the fitting functions,
integrals of the product
of the explicit functions of $\cos\theta$ and $\phi$ and the
$ae(\cos\theta,\phi)$:
\begin{eqnarray}
f_i(q_T,\phi) = \int_{0}^{\pi} g_i(\theta,\phi) ae(q_T,\cos\theta,\phi) d\cos\theta ,\\
i=-1,\ldots,7 \nonumber
\end{eqnarray}
where
\begin{eqnarray}
g_{-1}(\theta,\phi)&=&1+\cos^2{\theta},\;\;\;\;\;
g_0(\theta,\phi)=1/2(1-3\cos^2{\theta}) \nonumber \\
g_1(\theta,\phi)&=&\sin{2\theta} \cos{\phi},\;\;\,\,
g_2(\theta,\phi)=1/2\sin^2{\theta} \cos{2\phi} \nonumber \\
g_3(\theta,\phi)&=&\sin{\theta}\cos{\phi},\;\;\;\;\; 
g_4(\theta,\phi)=\cos{\theta} \nonumber \\
g_5(\theta,\phi)&=&\sin^2{\theta}\sin{2\phi},\;\;  
g_6(\theta,\phi)=\sin{2\theta} \sin{\phi} \nonumber \\
g_7(\theta,\phi)&=&\sin{\theta}\sin{\phi} \nonumber 
\end{eqnarray}
The $f_i$ functions can be calculated explicitly if we
know the acceptance and the efficiency of the detector.
Because we multiply by $ae(q_T,\cos\theta,\phi)$ before
integrating over $\cos\theta$, no $f_i$ is exactly zero.
\begin{figure}[!]
\includegraphics[scale=.44]{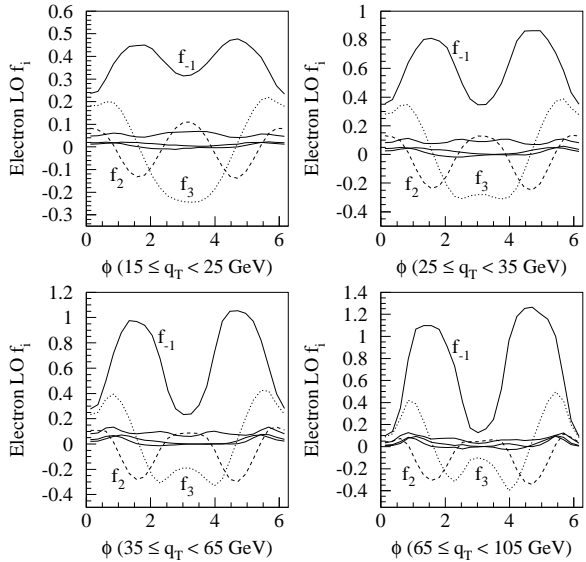}
\caption{The leading order $f_{-1,\cdots,4}(\phi)$.  These functions
are multiplied by the respective $W$ angular coefficients
to give us the experimentally observable $\phi$ distributions.}
\label{f105}
\end{figure}
\begin{figure}[!]
\includegraphics[scale=.44]{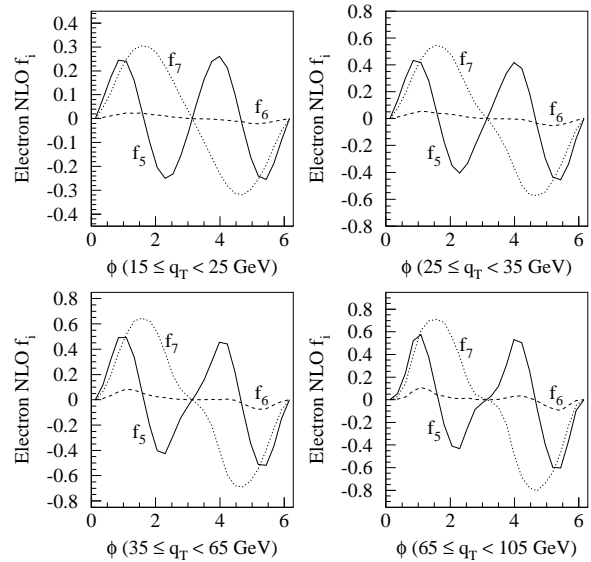}
\caption{The next-to-leading order $f_{5,6,7}(\phi)$.  These functions are
multiplied by the respective $W$ angular coefficients to give us
the next-to-leading order corrections to experimentally observable $\phi$ distributions.} \label{f10}
\end{figure}
\begin{figure}[!]
\includegraphics[scale=.44]{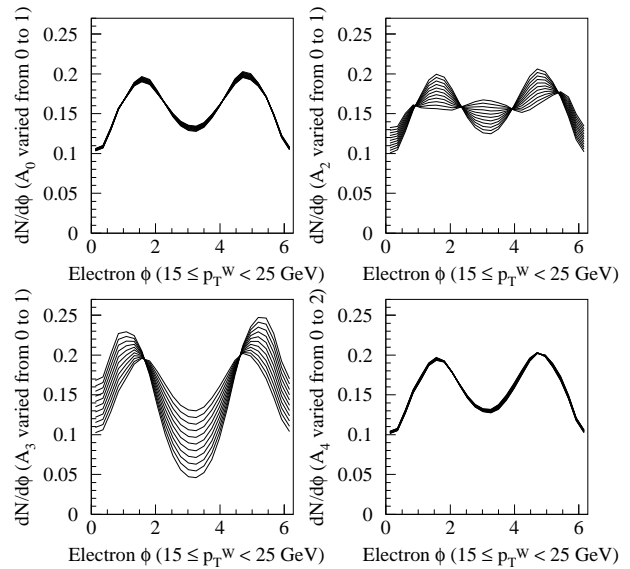}
\caption{The expected charged lepton $\phi$ distribution in the Collins-Soper $W$ rest-frame for the first $q_T$ bin, if we vary only one coefficient at the
time (with a step of 0.1) and keep the rest at the Standard Model value.  Only the $A_2$ and
$A_3$ significantly affect the shape of the distributions.  The same is true for the higher $q_T$ bins.}
\label{f9}
\end{figure}
\begin{figure}
\includegraphics[scale=.44]{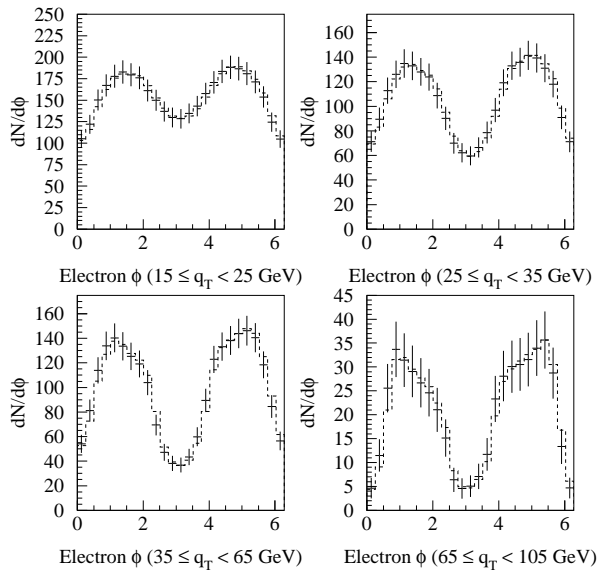}
\caption{The Monte Carlo prediction for the $\phi$ distribution,
including the detector acceptance and efficiency, for the four
$q_T$ bins (points). The size of the sample and error-bars
correspond to the expected yields at the Tevatron at
$\sqrt{s}=1.8$ TeV). Also shown is the result of the fit (dashed
line) using the theoretical prediction of Equation (\ref{new2})
and varying $A_2$ and $A_3$ simultaneously. The backgrounds are
ignored, only $W$+jet events are considered.} \label{f11}
\end{figure}
As a result, all coefficients are in principle measurable with the
$\phi$ analysis and not just $A_2$ and $A_3$, as Equation
(\ref{eqphi}) suggests. In practice, the $A_2$ and $A_3$ are
measurable with a greater statistical significance, because the
terms $A_if_i(\phi)$ are much smaller, for $i\ne2,3$.  As a result,
these terms affect less the $\phi$ distribution. 
Figures \ref{f105} and
\ref{f10} show the $f_i$ functions for electron acceptances and
efficiencies.  The shape of the $f_i$ functions is almost
identical for the muons. For perfect acceptance and no kinematic
cuts ($ae=1$), the only surviving $f_i$ functions would be
$f_{-1}$, $f_{2}$, $f_{3}$, $f_{5}$, and $f_{7}$, and they would
be equal to $8/3$, $2/3\cos 2\phi$, $\pi/2 \cos \phi$, $4/3 \sin
2\phi$ and $\pi/2 \sin 2\phi$, in accordance to Equation
(\ref{eqphi}). Figure \ref{f9} shows the $\phi$ distribution
(\ref{new2}) for the low $q_T$ bin,
with the background neglected and with only one 
coefficient varying at
a time.  We see that the $\phi$ distribution is primarily
sensitive to $A_2$ and $A_3$, and these coefficients are the easier
measurable ones with the $\phi$ analysis.
Figure \ref{f11} shows the Monte Carlo expected experimental
$\phi$ distributions for a data sample of the size of the
Tevatron Run I.

\begin{figure}[t]
\vspace{-0.38cm}
\includegraphics[scale=.44]{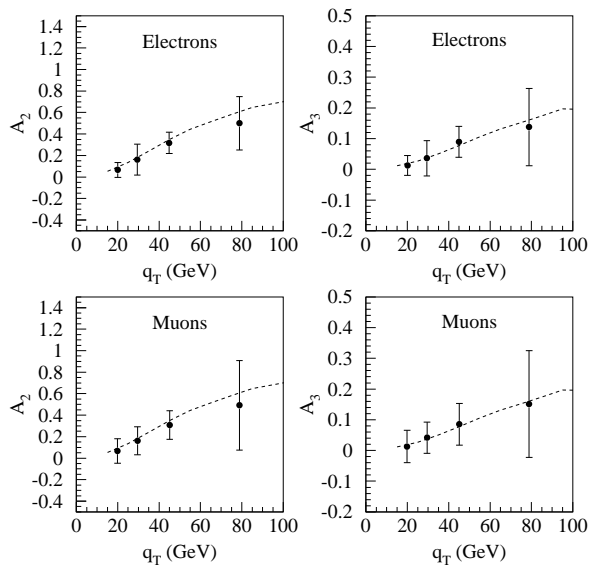}
\caption{The extracted angular coefficients, using the Monte Carlo
data of Figure \ref{f11} and fitting using the $f_i$ functions.
The angular coefficients that do not affect the shape of the
$\phi$ distribution are kept at their Standard Model values.
The errorbars are statistical only.}
\label{f12}
\end{figure}

The final step is to extract the angular coefficients using the
data of Figure \ref{f11}.  We keep the $A_{i\neq 2,3}$
coefficients frozen at their Standard Model values we determined
above and we fit the distributions to the $f_i$ varying $A_2$ and
$A_3$ simultaneously. The result of the fit can be seen in Figure
\ref{f11} and the extracted coefficients in Figure \ref{f12}. We
conclude that the measured angular coefficients are close to the
values we extracted in section \ref{sec:2}, verifying that the
method is self-consistent and could be used for an experimental
measurement of the $W$ angular coefficients.  The same technique
can be applied in $Z$ boson experimental studies -- which do not
demonstrate any problems in the kinematic reconstruction of the boson --
using the future statistically significant datasets of the Tevatron and the LHC.  

\section{Summary}
The Standard Model prediction for the angular coefficients
and the associated helicity cross sections of the $W$ production
in a hadron collider up to order $\alpha_s^2$ in QCD and
at $\sqrt{s}=1.8$ TeV was presented.
The experimental measurement of the angular distributions
is distorted due to the acceptances and efficiencies of the
detector and the
\vspace{11cm}

\hspace{-0.35cm}application of quality cuts to reduce backgrounds.
Two additional issues are the $W$ mass width effect and the
resolution of the two-fold ambiguity in the longitudinal
momentum of the neutrino.
We presented the effect of these factors on the angular
distributions and noted that both problems
do not affect the azimuthal angle of the charged lepton
in the CS frame.  Finally, we suggested a method of extracting
the angular coefficients without having to divide the
experimental data by Monte Carlo distributions of isotropic
$W$ decays.  Passing the generator data through a
detector simulator and analyzing the resulting data,
we were able to get back the angular coefficients
we determined from the direct analysis of
the generator data, demonstrating that this procedure is reliable
for the experimental measurement of the angular coefficients.

\begin{acknowledgments}
We wish to thank W. Giele for the interesting discussions and the
DYRAD generator support.  We also thank U. Baur and T. Junk
for reading the paper.
\end{acknowledgments}

\end{document}